\documentclass{aa}  

\usepackage{graphicx}
\usepackage[]{natbib}
\usepackage{txfonts}
\begin{document}

   \title{Magnetised Accretion Discs in Kerr Spacetimes II: Hot Spots}
   \author{Federico Garc\'{\i}a \inst{1,2}$^\star$, Ignacio F. Ranea-Sandoval
          \inst{3}\thanks{Fellow of CONICET, Argentina.}, \and  Tim Johannsen\inst{4,5}     
           }
\authorrunning{Garc\'ia et al.}
\institute{Instituto Argentino de Radioastronom\'{\i}a (CCT La Plata, CONICET), C.C.5, (1894) Villa Elisa, Buenos Aires, Argentina.
\and
Facultad de Ciencias Astron\'omicas y Geof\'{\i}sicas, Universidad Nacional de La Plata, Paseo del Bosque, B1900FWA La Plata, Argentina. \email{fgarcia@iar-conicet.gov.ar}
\and
Grupo de Gravitaci\'on, Astrof\'{\i}sica y Cosmolog\'{\i}a, Facultad de Ciencias Astron\'omicas y Geof\'{\i}sicas, Universidad Nacional de La Plata, Paseo del Bosque, B1900FWA La Plata, Argentina. 
\email{iranea@fcaglp.unlp.edu.ar}
\and
Perimeter Institute for Theoretical Physics, Waterloo, Ontario, N2L 2Y5, Canada. \email{tjohannsen@pitp.ca}
\and
Department of Physics and Astronomy, University of Waterloo, Waterloo, Ontario, N2L 3G1, Canada. 
}

   \date{Received ; accepted }


  \abstract
   {Quasi-periodic variability has been observed in a number of X-ray binaries harboring black hole candidates. In general relativity, black holes are uniquely described by the Kerr metric and, according to the cosmic censorship conjecture, curvature singularities always have to be clothed by an event horizon.}
   {In this paper, we study the effect of an external magnetic field on the observed light curves of orbiting hot spots in thin accretion discs around Kerr black holes and naked singularities.}
   {We employ a ray-tracing algorithm to calculate the light curves and power spectra of such hot spots as seen by a distant observer for uniform and dipolar magnetic field configurations assuming a weak coupling between the magnetic field and the disc matter.}
   {We show that the presence of an external dipolar magnetic field leads to potentially observable modifications of these signals for both Kerr black holes and naked singularities, while an external uniform magnetic field has practically no effect. In particular, we demonstrate that the emission from a hot spot orbiting near the innermost stable circular orbit of a naked singularity in a dipolar magnetic field can be significantly harder than the emission of the same hot spot in the absence of such a magnetic field.}
   {The comparison of our model with observational data may allow us study the geometry of magnetic fields around compact objects and to test the cosmic censorship conjecture in conjunction with other observables such as thermal continuum spectra and iron line profiles.}


   \maketitle
%

\section{Introduction}

Understanding the final fate of an initial mass distribution after its gravitational collapse is still one of the most active fields of research within the general relativistic community. Although general relativity predicts the appearance of spacetime singularities, the formation of an event horizon covering them is not necessarily required. 

Back in 1969, Roger Penrose proposed the cosmic censorship conjecture~\citep[CCC; see][for a review on the subject]{CCC}. The weak form of the CCC states, in simple language, that in the final state of the gravitational collapse of ``normal'' matter each spacetime singularity must be hidden behind an event horizon. Whether this conjecture is valid or not is amongst the most important open problems in the classical theory of general relativity. Although great efforts -- following different lines of thought -- have been made over the past 40 years, there is still no definitive answer to the question whether the (weak and strong) CCC are valid or not \citep[see][and references therein]{joshi}. 

To date, there is still no conclusive observational evidence of the actual nature of compact objects and no direct evidence of an event horizon has been found yet \citep[but see, e.g.,][]{Broderick09}. For this reason, finding observable features that could help to distinguish between black holes (BHs) and naked singularities (NSs) should be considered relevant as they could shed some light on the validity of the CCC.

The study of potentially observable differences between accretion discs around BHs and NSs is important as it may give us an observational tool to determine the nature of the central object. Circular equatorial orbits of particles comprising the disc are a key ingredient to the study of more realistic disc models. \citet{Stuchlik} studied circular equatorial orbits in the spacetime of a rotating NS in detail. A recent comparative study of properties of accretion discs around BHs and NSs can be found in \citet{kovacs} and references therein.

The physics of accretion discs is governed by a combination of different complex processes: gravitomagnetohydrodynamics, turbulent viscous fluids, and radiation fields, among others. The theoretical interpretation of observations of compact X-ray sources is extremely relevant to the understanding of compact-object stellar astrophysics \citep[see, e.g.,][]{2010csxs.book.....L}. Such studies reveal that magnetic fields are usually associated with accretion discs around compact objects. Understanding this connection could serve as a tool to provide information related to the nature of the central compact object and to discard controversial theoretical models such as NSs.

The presence of magnetic fields affects the trajectories of particles orbiting around a rotating compact object and changes the location of the innermost stable circular orbit \citep[ISCO;][]{pv, wiita,iyer.et.al,rsv}. Such modifications may give rise to observable quantities that could be used to distinguish between different theoretical models for compact objects. 

Recently, \citet{RSG2015a} proposed a model of magnetised accretion discs formed in generalised Kerr spacetimes. The model is based on the analysis of the circular equatorial trajectories followed by charged particles in a Kerr spacetime under the presence of an external magnetic field, previously developed by \citet{rsv}. In their paper, the authors study the simplest exact solutions to the Maxwell equations in a Kerr background obtained by \citet{pttson}, introducing two particular magnetic field configurations: uniform and dipolar.

Even though there are studies that predict the linear instability of completely collapsed objects without event horizons, such as NSs (see, e.g., \citealp{dottietal21} and \citealp{dottietal23} for studies related to the Kerr solution and \citealp{sspinar1} and \citealp{sspinar2} for superspinars), through this work, we study potentially observable differences in the light curves produced by a hot spot orbiting at the innermost regions of an accretion disc formed around both BHs and NSs, with the idea of developing a tool to probe the nature of the spacetime in the strong gravitational regime close to fully-collapsed objects.

We extend the model presented in \citet{RSG2015a} based on an adapted version of the free source code YNOGK \citep{ynogk} as well as on the ray-tracing code developed by \citet{j-psaltis}. This version allows us to consider Kerr NSs and to include magnetic field effects on the trajectories followed by particles with an effective electric charge. Using our machinery, we model the observed light curves of hot spots formed in magnetised accretion discs in a Kerr background and analyse their power spectra. The developed tool provides the possibility to perform an analysis that is independent of the thermal continuum spectra and broad iron-$K_\alpha$ emission line methods, which have proved very useful for measuring the spin parameters of a number of compact objects assuming a Kerr background. See \citet{obs_a} and \citet{obs_b} for respective reviews. 

Another observational characteristic of the X-ray emission from BH candidates is its high-frequency variability, detected in their power spectra, known as quasi-periodic oscillations \citep[QPOs;][]{qpos1,qpos2}. This particular feature is often observed at commensurate integer ratios in accreting systems with a central compact object in the case of high-frequency QPOs \citep{miller2001,remillard2002}. Such observations are a promising probe of the innermost part of the accretion disc and of the nature of the compact object itself \citep[see, e.g.,][]{lamb2003,psaltis2004}. Further, they can be used as a tool to infer the spin parameter of the compact object \citep[e.g.,][]{Peres97,HS1998,HS1999,Silbergleit01,Kluzniak01,Abramowicz03}.

The comparison between observational data from X-ray binary systems and our theoretical model might be a useful tool to get insight into the physical properties related to the dark massive compact objects present in such binary systems and also to test Penrose's conjecture in astrophysical environments. Moreover, as we are taking into account the effect of magnetic fields on observable quantities, this work represents a new tool to estimate the strength and geometry of magnetic fields in the neighborhood of an accretion disc.

The work is organised as follows: in Section \ref{mag-disc}, we present the basics of magnetised accretion discs and hot spots as well as a description of the theoretical framework used. In Section \ref{num-code}, we describe our numerical code. Section \ref{results} is devoted to the presentation of our results for light curves and power spectra for corotating hot spots in accretion discs formed in a generic Kerr background spacetime with an external (uniform or dipolar) magnetic field. 
Finally, we discuss our results and present conclusions in Section \ref{conclusions}.

\section{The model} \label{mag-disc}

\citet{sys} and \citet{pyt} set the grounds for studying geometrically thin, optically thick accretion discs in astrophysical environments.

Magnetic fields play a decisive role regarding gas dynamics in jet physics and accretion processes into compact objects. Page-Thorne accretion discs are formed by matter particles orbiting around the central object following time-like circular equatorial geodesics. In order to model magnetised accretion discs, we incorporate the corrections to circular trajectories followed by charged particles in a Kerr spacetime with an external global magnetic field \citep{rsv}. 

Like other accretion discs models, the one proposed by \citet{RSG2015a} is not completely self consistent. Matter in the disc and the magnetic field do not alter the spacetime geometry. Moreover, the magnetic field structure is not tied to the plasma motion. For these reasons, we only consider magnetic fields of small strengths weakly interacting with the plasma in this paper.

\subsection{The Kerr spacetime}

A completely collapsed object with mass $M$ and angular momentum $J\equiv a M$ is described by the vacuum solution to the Einstein field equations obtained by \citet{kerr}:
\begin{eqnarray}
ds^2  &=& \frac{\Delta-a^2 \sin ^2 \theta}{\Sigma}dt^2 + 2 a M \sin ^2 \theta\frac{r}{\Sigma}dtd\phi - \frac{\Sigma}{\Delta}dr^2 \nonumber \\
&&{} - \Sigma d\theta ^2 - 
\frac{(r^2+a^2)^2-\Delta a^2 \sin ^2\theta}{\Sigma}\sin ^2\theta d\phi ^2 , \label{kerrbl}
\end{eqnarray}
where $\Delta \equiv r^2-2Mr+a^2$, $\Sigma \equiv r^2+a^2 \cos ^2 \theta$. Here, we use Boyer-Lindquist coordinates and $(+,-,-,-)$ metric signature. In this section, we also use geometric units and set $c=G=1$, where $c$ and $G$ are the speed of light and the gravitational constant, respectively.

In the sub-extreme case ($0\leq |a|<M$), the Kerr spacetime harbors a BH with two horizons located at the roots of the function $\Delta$\footnote{In the Schwarzschild case ($a=0$), the inner horizon merges with the central singularity located at the radius $r=0$.}. In the extreme case ($|a|=M$), the two horizons merge as the root of $\Delta$ becomes a double root. In both cases, the outer horizon causally disconnects the interior domain from the external universe hiding the curvature singularity. In the super-extreme case ($|a|>M$), the ring singularity is causally connected to future null infinity as no horizons are present, because the roots of the function $\Delta$ become complex. In this case, the spacetime harbors what is usually called a Kerr NS. Note, however, that causality is violated everywhere around a Kerr NS, because any event in that spacetime can be connected to any other event by both a future and a past directed timelike curve \citep{Carter68,Carter73}.

\subsection{Magnetised accretion discs}

Following \citet{RSG2015a} we incorporate two different magnetic field configurations: the simplest uniform (\rm{U}) case and a dipolar (\rm{D}) one, which could be generated through the Poynting-Robertson Cosmic Battery mechanism \citep{CB-1,CB-2}.

The four-vector potential in the uniform case can be expressed as \citep{pttson}
\begin{eqnarray}
A_{\mu}^{\rm U} &=& \left(A_t^{\rm U},0,0,A_\phi^{\rm U}\right),
\end{eqnarray}
where
\begin{eqnarray}
A_t^{\rm U} &=& -aB\left(1-\frac{Mr}{\Sigma}\left(2- \sin ^2 \theta\right)\right), \nonumber \\
A_\phi^{\rm U}  &=& \frac{B \sin ^2 \theta}{2\Sigma} \left[\left(r^2+a^2\right)^2 - \Delta a^2  \sin ^2 \theta -4Ma^2r\right]
\end{eqnarray}
and where $B$ is the magnetic field strength. For this magnetic field configuration, the coupling constant between the effective charge and the magnetic field strength is defined as
\begin{equation}
\lambda _{\rm U} \equiv \frac{e}{m} B M,
\end{equation}
where $e$ and $m$ are the charge and rest mass of the particles, respectively.

For the dipolar case \citep{pttson}, the non-zero components of the four-vector potential 
\begin{eqnarray}
A_{\mu}^{\rm D} &=& \left(A_t^{\rm D},0,0,A_\phi^{\rm D}\right)
\end{eqnarray}
can be written as
\begin{eqnarray}
A_t^{\rm D} &=& -\frac{3a\mu}{2 \left(1-a^2\right) \Sigma}\left(f_1\left(r,\theta\right) \frac{\ln \Theta}{2 \sqrt{1-a^2}}
-\left(r-M \cos ^2 \theta \right)\right), \nonumber \\
A_\phi^{\rm D} &=& -\frac{3\mu \sin ^2 \theta}{4 \left(1-a^2\right) \Sigma} \left(f_2\left(r,\theta\right) +f_3\left(r,\theta\right)\frac{\ln \Theta }{2 \sqrt{1-a^2}} \right), 
\end{eqnarray}
where $\mu$ is the magnetic dipole moment (assumed to be aligned with the spin axis), $f_i\left(r,\theta\right), i=1,2,3$, are functions defined as
\begin{eqnarray}
f_1\left(r,\theta\right) &\equiv& r\left(r-M\right)+\left(a^2-Mr\right)\cos ^2 \theta , \nonumber \\
f_2\left(r,\theta\right) &\equiv& \left(r-M\right)a^2\cos ^2 \theta + r\left(r^2+Mr+2a^2\right) , \nonumber \\
f_3\left(r,\theta\right) &\equiv& - \left[r\left(r^3+ a^2r-2Ma^2\right)+\Delta a^2 \cos ^2 \theta \right],
\end{eqnarray}
and the function $\Theta$ is given by
\begin{eqnarray}
\Theta &=& \frac{r-M+\sqrt{1-a^2}}{r-M-\sqrt{1-a^2}}.
\end{eqnarray}

For this magnetic field configuration, the coupling constant between the effective charge and the magnetic field strength is

\begin{equation}
\lambda _{\rm D} \equiv \frac{e \mu}{M^2}. 
\end{equation}

\subsection{Emission from a hot spot}

The simplest model of a hot spot was developed by \citet{HS1998,HS1999} and assumes that isotropic, monochromatic emission originates from a small region that corotates with the accretion disc following a circular trajectory. In this ``blob'' model, the emissivity is assumed to be a Gaussian distribution of the distance from the centre of the hot spot:
\begin{equation} \label{emissivity}
\epsilon (\mathbf{x}) \propto {\rm exp} \left(-\frac{|\mathbf{x}-\mathbf{x}_{\rm hs}(t)|^2}{2R^2_{\rm hs}} \right).
\end{equation}

Here, the spatial 3-vector, $\mathbf{x}$, is given in pseudo-Cartesian coordinates for the Kerr spacetime. The size, $R_{\rm hs}$, of the hot spot has to be small compared to the radius of its centre. If this were not the case, as the model assumes that all points belonging to the hot spot have the same 4-velocity of the ``guiding'' trajectory of the centre, tidal effects would eventually destroy the hot spot. 

In order to properly compute the observed light curve it is essential to take into account the time delay effects in Kerr spacetime. After obtaining the hot spot trajectory parametrised via the coordinate time $t$, a ray-tracing map between the disc and the observer must be used to obtain the observed flux as a function of time. For each photon bundle intersection point, one can calculate a position-dependent time delay, $\Delta t$. This means that in order to calculate the associated observer time, $t_{\rm obs}$, first it is necessary to determine the position of the hot spot at emission time, $t_{\rm em} = t_{\rm obs} - \Delta t$, which is different for each point of the accretion disc. 

Repeating this procedure along the whole rotation period results in a time-dependent spectrum or spectrogram. After integration over frequencies the light curve is obtained. In this paper, we will set the arbitrary phase to the value $\phi=0.5$ at the time when the emission peaks for each light curve.

The flux from a hot spot experiences variations due to two fundamental processes as the hot spot is orbiting around the compact object. Frequency shifts of the emitted photons as well as flux amplifications and deamplifications occur due to Doppler boosting, gravitational lensing, and relativistic beaming.

\section{The numerical code} \label{num-code}

\begin{figure*}
\centering
{\tiny $a=0.5 \hspace{0.45\textwidth} a=2.0$}\\
\vspace{0.1cm}
\includegraphics[width=0.45\textwidth]{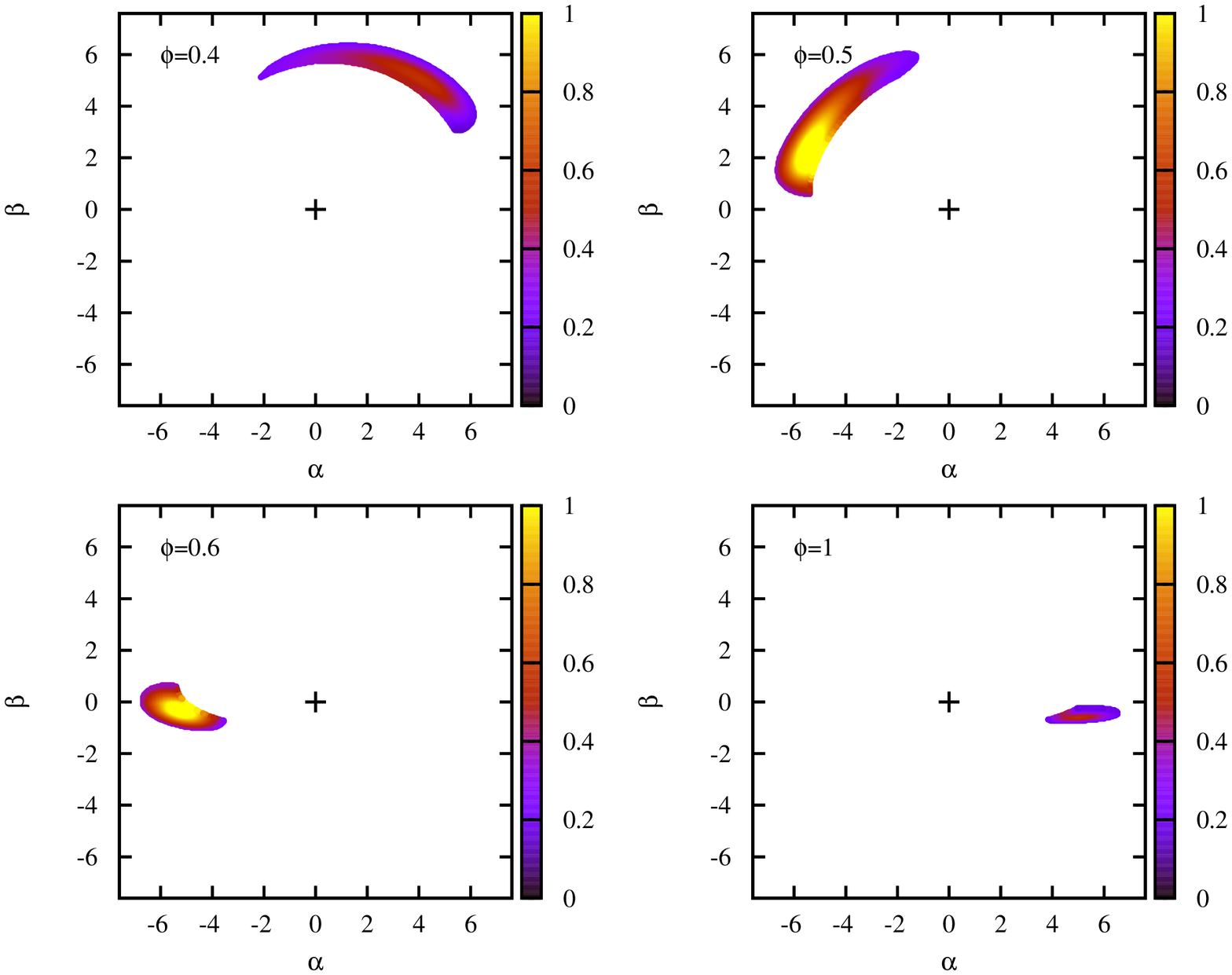} \hspace{1.2cm}
\includegraphics[width=0.45\textwidth]{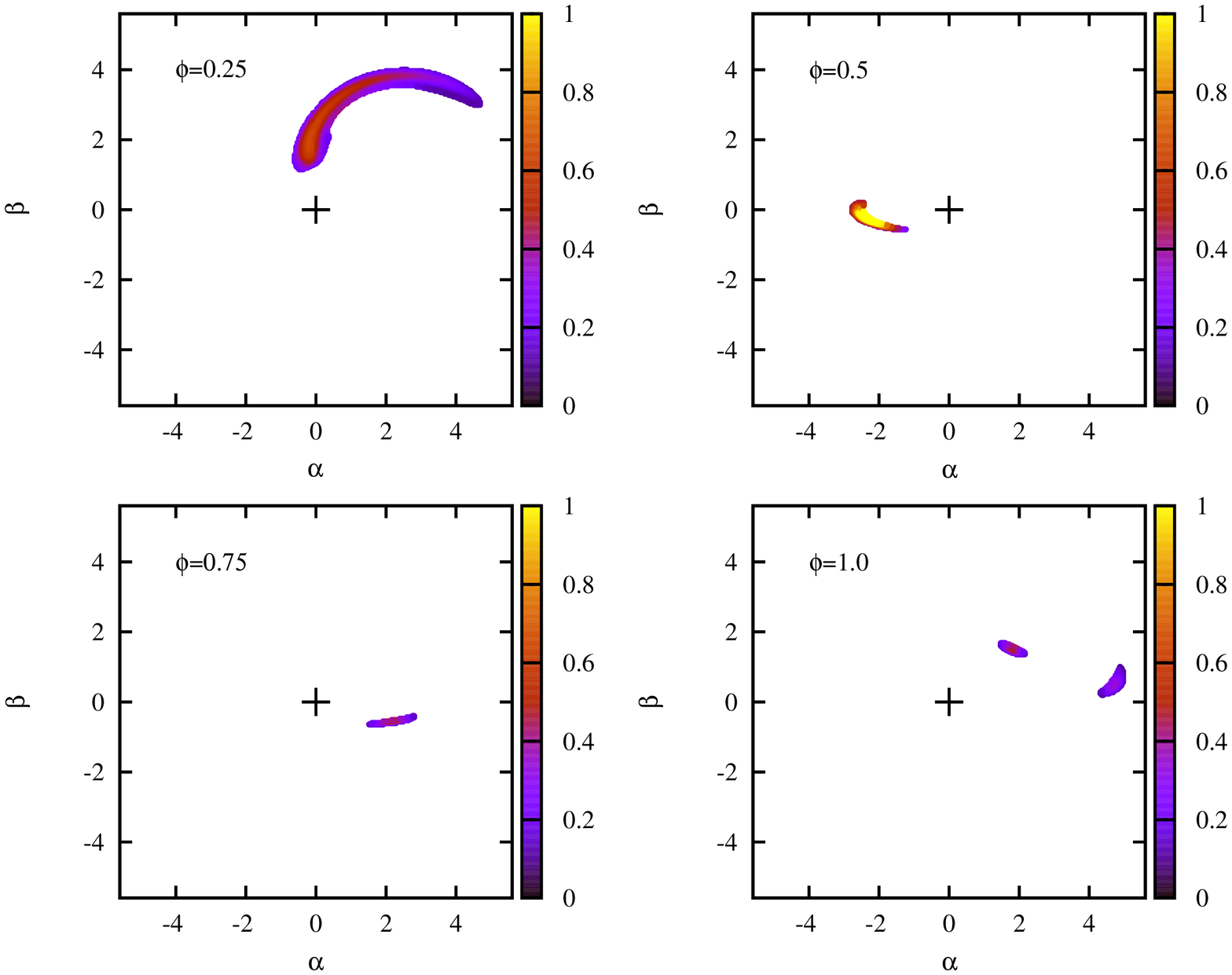}
 \caption{Snapshots of the motion of a hot spot formed near the ISCO of a black hole with a spin $a=0.5r_g$ (left panels) and a naked singularity with a spin $a=2r_g$ (right panels), both observed at an inclination of $i=75^\circ$ at four different phases $\phi$. The colour bar indicates the observed flux in arbitrary units. The impact parameters $\alpha$ and $\beta$ are expressed in gravitational radii.}
 \label{fig0}
\end{figure*}

In this section we present how our numerical code operates to obtain the light curve of a corotating hot spot in a magnetised accretion disc that generalises the classical model \citet{pyt} model.

In order to obtain the light curves as seen by a distant observer at an arbitrary position, we incorporate a ray-tracing technique to calculate the geodesics of photons between a plane placed at the position of the observer and the surface of the disc. For this purpose, we adapted both the public semi-analytic code YNOGK \citep{ynogk} and the fully-numerical code from \citet{j-psaltis} to incorporate the orbits corresponding to circular trajectories followed by charged particles coupled to a magnetised Kerr spacetime through different values of the $\lambda_{\rm U,D}$ parameters, generalising the equations to allow for the appearance of NSs for $|a|>M$. With this machinery we proceed as follows:

\begin{itemize}
\item We calculate the properties of circular trajectories followed by
matter in the disc at radii $r_{\rm hs}$.
\item We evaluate the orbital velocity at the hot spot radius to define the phases.
\item For each discrete phase, we obtain the image of the hot spot and we evaluate the relativistic effects experienced by photons emitted from disc as they travel to the observer by means of the ray-tracing code.
\end{itemize}

For the calculations presented in this paper, phase and energy spaces are discretised to 100 steps, and snapshot images are constructed using  a $500 \times 500$ pixel resolution. We also fixed $r_{\rm hs}=1.1 r_{\rm ISCO}$ and $R_{\rm hs}= 0.1 r_{\rm hs}$ as we are only interested in the imprints of the motion of a compact hot spot at the innermost region of the accretion disc.

In Fig.~\ref{fig0} we show selected snapshots of the observed flux of a hot spot located at the innermost part of a non-magnetised accretion disc formed around a BH with spin parameter $a=0.5r_g$ (left panels) and a NS with spin parameter $a=2r_g$ (right panels), as seen by a distant observer at an inclination angle of $75^\circ$. The different colours indicate the observed flux in arbitrary units and $\alpha$ and $\beta$ are coordinates in the observer's plane expressed in gravitational radii $r_g\equiv GM/c^2$.

A clear difference between the two opposite cases can be observed at phase $\phi = 1$ where, in the case of the NS, two different images of the same hot spot can be observed (due to light-bending effects and the absence of an event horizon). This secondary image arises because the corresponding photons make one full revolution around the NS. A similar feature was observed in the paper of \citet{li-bambi-2014}. In the BH case, the emission of the hot spot is negligible for $\sim 70\%$ of the time, becoming important only in the 0.35--0.65 phase range. Thus we selected snapshots at $\phi$= 0.4, 0.5, 0.6 and 1.0. On the contrary, in the NS case, the hot spot is brighter for roughly half of the period and thus we chose equidistant phases $\phi$= 0.25, 0.5, 0.75 and 1.0 to illustrate the snapshots.

The strong elongation of the observed image of the hot spot at $\phi \sim 0.5$ (for the BH) and $\phi \sim 0.25$ (for the NS) is a direct consequence of relativistic time dilation. As we mentioned, the phase is arbitrarily chosen so that the observed emission peaks at $\phi=0.5$. In both the BH and NS cases, this phase corresponds approximately to the moment when the hot spot velocity points in the direction to the observer, as a consequence of Doppler boosting.

\section{Results} \label{results}

After presenting the central aspects of our model and numerical code, in what follows we center the attention on our results and their implications. We divide them in two subsections: one for the light curves and one for the power spectra.

\subsection{Light curves of hot spots in magnetised accretion discs}

\begin{figure*}
\centering
\includegraphics[height=0.9\textwidth, angle=-90]{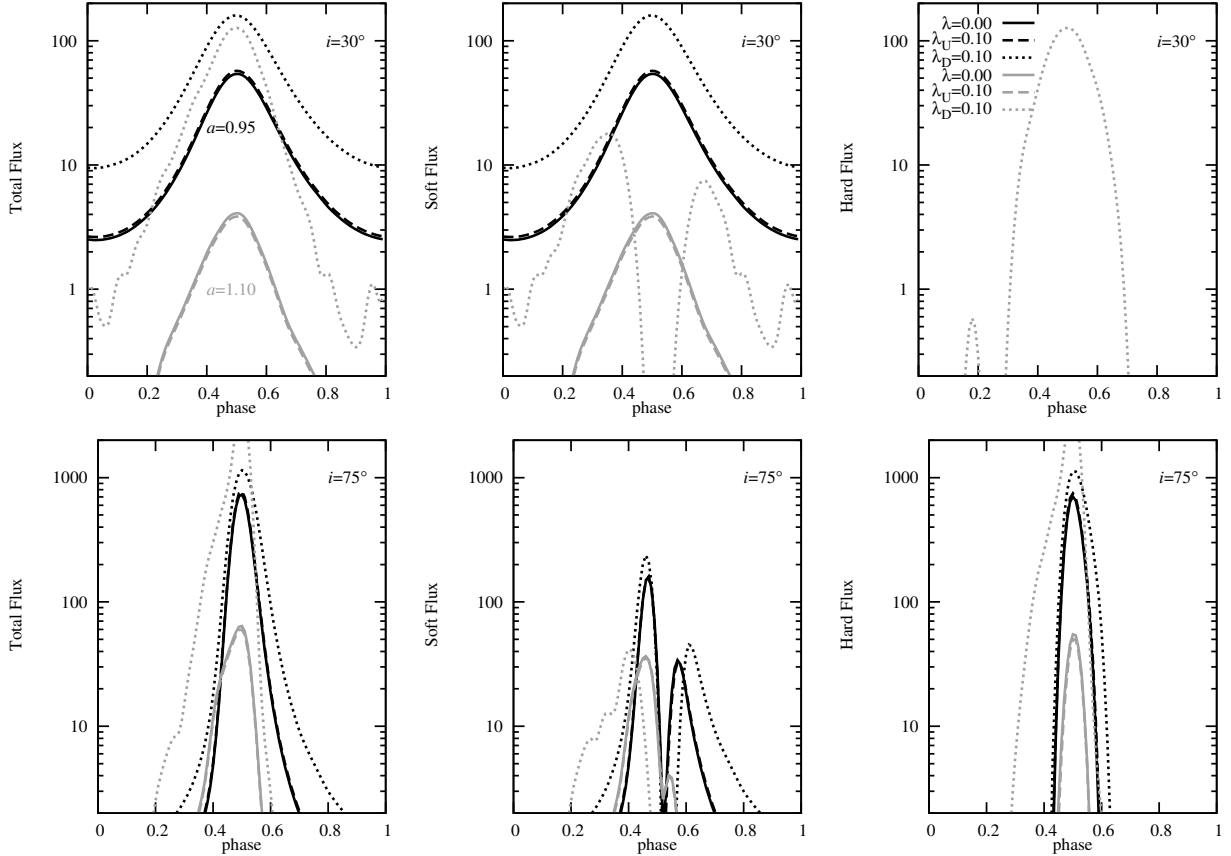}
 \caption{Light curves of corotating hot spots at the innermost part of accretion discs around a BH with rotational parameter $a=0.95r_g$ (black lines) and a NS with $a=1.10r_g$ (gray lines) viewed at inclinations of $i=30^\circ$ (top panels) and $i=75^\circ$ (bottom panels), respectively. Solid lines are used for non-magnetised accretion discs and dashed and dotted lines correspond to uniform and dipolar magnetised cases, respectively. The left panels show the total light curves which are fainter in the case of NSs. The center and right panels show the same light curves, separated into soft and hard energy bands with redshifts $g\leq0.9$ and $g>0.9$, respectively. A dipolar magnetic field configuration leads to noticeable changes in the shapes and amplitudes of these light curves, while a uniform magnetic field configuration has practically no effect on the light curves.}
 \label{fig1}
\end{figure*}

Implementing the numerical algorithm described in the previous section,
we have been able to generate a family of theoretical light curves emitted by corotating hot spots formed in magnetised accretion discs around both BHs and NSs. 

In Fig.~\ref{fig1} we show the light curves of hot spots formed in the innermost part of magnetised accretion discs. In the left panels, we show the total flux as a function of the phase angle. The top and bottom panels correspond to inclinations of $i=30^\circ$ and $i=75^\circ$, respectively. Clear differences can be seen between those formed around $a=0.95r_g$ BHs (black lines) and $a=1.10r_g$ NSs (gray lines) which are fainter in the latter case. Moreover, narrower (broader) light curves are found for high (low) inclinations. The presence of a uniform magnetic field practically does not affect the light curves, while the presence of a dipolar one strongly changes their shapes, as it is larger at the innermost parts of the disc.

When the observed photons are separated into soft ($g\leq0.9$) and hard ($g>0.9$) energy bands (central and right panels, respectively), where $g$ is the photon redshift, the light curves can have a very different shape. At high inclinations, in the case of either a BH or a NS as well as at low inclinations in the case of a NS in the presence of a dipolar magnetic field, the light curve has two peaks in the soft energy band, one before and one after the phase corresponding to the peak of the total light curve. These peaks are an interesting characteristic for a determination of the parameter $\lambda_{\rm D}$. 

In the hard energy band, the contribution of the hard flux to the total observed flux is negligible for low inclinations except for the case of a dipolar magnetic field around NSs (see top right panel of Fig.~\ref{fig1}). The explanation for this phenomenon lies on the fact that the ISCO is closer to the curvature singularity than in the non-magnetised case \citep{RSG2015a} causing the effect of Doppler boosting to be sufficiently strong to overcome the effect of the gravitational redshift.

The origin of the wiggle and second bump present in the top panels of Fig.~\ref{fig1} for the $a=1.10r_g$ NS with $\lambda _{\rm D} = 0.1$ can be tracked down to the appearance of the secondary image (see right bottom panels of Fig.~\ref{fig0}). This particular (and potentially observable) feature could be used as a tool to distinguish the nature of the central compact object.

\subsection{Power spectra and quasi-periodic oscillations}

\begin{table*}[]
\centering
\caption{Innermost stable circular orbits and orbital frequencies associated with the hot spots}
\label{table}
\begin{tabular}{cc | cc cc cc cc}
\hline
\hline

             &                       & \multicolumn{2}{c}{$a=0.5r_g$}                & \multicolumn{2}{c}{$a=0.95r_g$}               & \multicolumn{2}{c}{$a=1.1r_g$}                & \multicolumn{2}{c}{$a=2r_g$}                  \\
$M$          & $\lambda$             & $r_{\rm isco}$ {[}km{]} & $\nu_0$ {[}Hz{]} & $r_{\rm isco}$ {[}km{]} & $\nu_0$ {[}Hz{]} & $r_{\rm isco}$ {[}km{]} & $\nu_0$ {[}Hz{]} & $r_{\rm isco}$ {[}km{]} & $\nu_0$ {[}Hz{]} \\

\hline

             & $\lambda=0$           & 31.37                   & 702.5            & 14.36                   & 1765.9           & 4.94                    & 3918.6           & 9.36                    & 1881.9           \\
$5M_\odot$  & $\lambda_{\rm U}=0.1$ & 31.37                   & 528.5            & 14.36                   & 1720.8           & 4.94                    & 3944.3           & 9.36                    & 1914.2           \\
             & $\lambda_{\rm D}=0.1$ & 32.60                   & 663.8            & 18.28                   & 1340.6           & 4.19                    & 2365.3           & 9.03                    & 1862.6           \\
            
            \hline
             & $\lambda=0$           & 62.74                   & 351.3            & 28.71                   & 883.1            & 9.89                    & 1959.6           & 18.72                   & 941.1            \\
$10M_\odot$ & $\lambda_{\rm U}=0.1$ & 62.74                   & 264.3            & 28.71                   & 860.5            & 9.89                    & 1972.5           & 18.72                   & 957.2            \\
             & $\lambda_{\rm D}=0.1$ & 65.20                   & 332.0            & 36.57                   & 670.4            & 8.37                    & 1182.8           & 18.07                   & 931.4           \\

\hline

\end{tabular}
\end{table*}

\begin{figure*}
\centering
\includegraphics[height=0.9\textwidth, angle=-90]{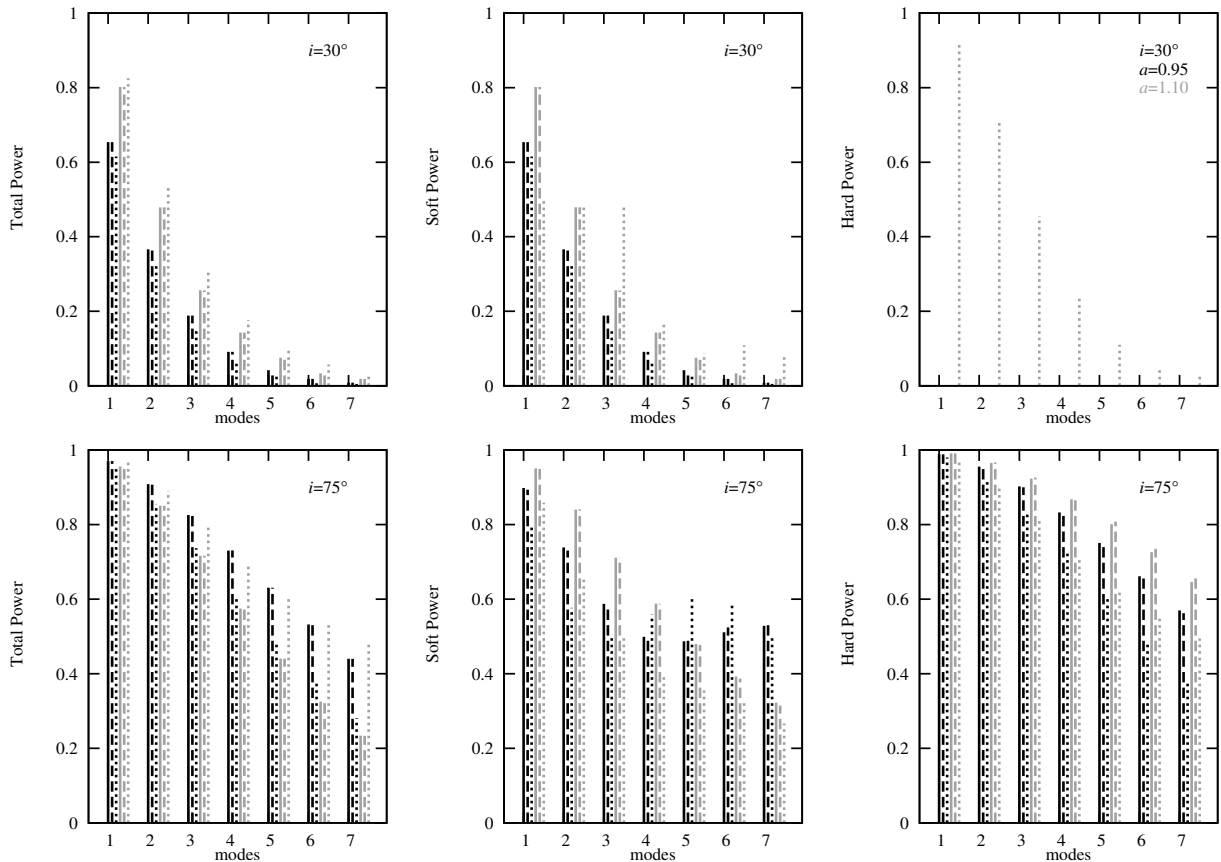}
 \caption{Fundamental modes and lowest-order harmonics of the light curves presented in Fig.~\ref{fig1}. The number of potentially observable modes and their relative amplitudes are strongly dependent on the type of the compact object, the inclination angle $i$, and the presence of a dipolar magnetic field. At small inclinations, the modes in the hard energy band are only visible in the NS case in a magnetised accretion disc with $\lambda_{\rm D} = 0.1$, which makes this a particularly interesting tool for a determination of the parameter $\lambda_{\rm D}$.}
 \label{fig2}
\end{figure*}

Computing the Fast-Fourier Transforms (FFTs) of the simulated light curves allows us to obtain the corresponding signals in the frequency domain. These frequencies can, in principle, be compared with the high-frequency X-ray variability seen in the observational data from X-ray binary systems. Here, we focus on the fundamental mode and higher-order harmonics of each light curve. In Table~\ref{table} we show the locations of the ISCO and the corresponding fundamental modes of the orbiting hot spot for BHs with spins $a=0.5r_g$ and $0.95r_g$ as well as for NSs with spins $a=1.10r_g$ and $2.0r_g$, assuming typical masses of $5M_\odot$ and $10M_\odot$ for each case.

In Fig.~\ref{fig2} we present FFTs of the light curves shown in Fig.~\ref{fig1}. The mode patterns are strongly dependent on the inclination angle, $i$, both in the number of potentially observable modes and in their relative amplitudes. Differences for large inclinations (bottom panels) between the FFTs of the light curves in the soft and hard energy bands are evident. For small inclinations, FFTs in the hard energy band are only visible in the NS case in a magnetised accretion disc with $\lambda_{\rm D} = 0.1$, which makes this an interesting tool for a determination of the parameter $\lambda_{\rm D}$. For large inclinations, the situation changes and modes are visible in both the soft and hard energy bands. The variations between magnetised and non-magnetised FFTs are too small to be detectable.

\section{Discussion} \label{conclusions}

One of the main problems in general relativity and high-energy astrophysics is related to the question whether the CCC is valid or not. Astrophysical studies of compact objects are highly important as they can provide a unique tool to test the validity of the weak CCC. In this sense, the development of theoretical tools that can be used to determine the nature of the compact object in an X-ray binary system has been an area of great activity during the last decades. 

The nature and strength of the magnetic field in the region very close to the compact object present in these type of binaries is of big interest and could help us to better understand the complex scenario of the formation of accretion discs. As currently most of the estimates for the magnetic fields found around compact objects are obtained by means of the properties of the non-thermal radiation produced far away from the disc as well as of Faraday rotation measurements \citep[see, e.g.,][]{FRM,FRM2}, the study of the effects caused by an external magnetic field on the emission of hot spots of accretion discs becomes relevant for these purposes.

Motivated by these two reasons, we continued the work presented in \citet{RSG2015a} and developed a new tool that can be used to estimate the presence of a magnetic field in the neighborhood regions of an accretion disc formed around a Kerr compact object. In the framework of the magnetised Page-Thorne accretion disc model developed by \citet{RSG2015a} we studied non-magnetised accretion discs, recovering well known results from \citet{pyt}. We also extended these results by analysing the effects of an external uniform or dipolar magnetic field on the hot spot light curves and power spectra. For the first time we present light curves of hot spots formed in accretion discs around NSs.

We showed that the presence of an external dipolar magnetic field leads to potentially observable modifications of the light curves and power spectra for both Kerr BHs and NSs, while an external uniform magnetic field has practically no effect. In particular, we demonstrated that the emission from a hot spot orbiting near the ISCO of a NS in a dipolar magnetic field can be significantly harder than the emission of the same hot spot in the absence of such a magnetic field, at least for the spin values we considered.

Highly precise detections of fundamental modes and higher-order harmonics of QPOs in X-ray binaries will become possible with future X-ray missions such as the Large Observatory For x-ray Timing \citep[LOFT;][]{Feroci14}. Observations of QPOs could also provide a test of the Kerr nature of BH candidates and, thereby, of general relativity \citep{JP11,Bambi12}. A direct image of a handful of supermassive BHs and, hence, of an event horizon may be obtained by the Event Horizon Telescope \citep{Doele09a,Doele09b,Fish09}.

\begin{acknowledgements}
We would like to thank Dr. M.~M.~Miller Bertolami for borrowing us the computer with which most of the numerical calculations were performed for this paper. IFRS acknowledges support from Universidad Nacional de La Plata. IFRS and FG are fellows of CONICET. TJ was supported in part by Perimeter Institute for Theoretical Physics. Research at Perimeter Institute is supported by the Government of Canada through Industry Canada and by the Province of Ontario through the Ministry of Research and Innovation.
\end{acknowledgements}

\end{document}